\documentstyle[11pt,aaspp4]{article}
\begin{document}
\def\hh{\, h^{-1}}
\newcommand{\wth}{$w(\theta)$}
\newcommand{\Lya}{Ly$\alpha$}
\newcommand{\Lyb}{Lyman~$\beta$}
\newcommand{\Hb}{H$\beta$}
\newcommand{\msun}{M$_{\odot}$}
\newcommand{\sfr}{M$_{\odot}$ yr$^{-1}$}
\newcommand{\za}{$z_{\rm abs}$}
\newcommand{\ze}{$z_{\rm em}$}
\newcommand{\cmtwo}{cm$^{-2}$}
\newcommand{\nhi}{$N$(H$^0$)}
\newcommand{\degpoint}{\mbox{$^\circ\mskip-7.0mu.\,$}}
\newcommand{\halpha}{\mbox{H$\alpha$}}
\newcommand{\hbeta}{\mbox{H$\beta$}}
\newcommand{\hgamma}{\mbox{H$\gamma$}}
\newcommand{\kms}{\,km~s$^{-1}$}      
\newcommand{\minpoint}{\mbox{$'\mskip-4.7mu.\mskip0.8mu$}}
\newcommand{\mv}{\mbox{$m_{_V}$}}
\newcommand{\Mv}{\mbox{$M_{_V}$}}
\newcommand{\peryr}{\mbox{$\>\rm yr^{-1}$}}
\newcommand{\secpoint}{\mbox{$''\mskip-7.6mu.\,$}}
\newcommand{\sqdeg}{\mbox{${\rm deg}^2$}}
\newcommand{\squig}{\sim\!\!}
\newcommand{\subsun}{\mbox{$_{\twelvesy\odot}$}}
\newcommand{\et}{{\it et al.}~}
\newcommand{\er}[2]{$_{-#1}^{+#2}$}
\def\h50{\, h_{50}^{-1}}
\def\hbl{km$^{-1}$~Mpc$^{-1}$}
\def\ltsima{$\; \buildrel < \over \sim \;$}
\def\simlt{\lower.5ex\hbox{\ltsima}}
\def\gtsima{$\; \buildrel > \over \sim \;$}
\def\simgt{\lower.5ex\hbox{\gtsima}}
\def\arcs{$''~$}
\def\arcm{$'~$}
\title{THE ANGULAR CLUSTERING OF LYMAN-BREAK GALAXIES AT REDSHIFT $z\sim 3$
\altaffilmark{1}}
\author{\sc Mauro Giavalisco\altaffilmark{2}}
\affil{Observatories of the Carnegie Institution of Washington, 813 Santa
Barbara Street, Pasadena, CA 91101}
\affil{e-mail: mauro@ociw.edu}
\author{\sc Charles C. Steidel\altaffilmark{3,4} and Kurt L. Adelberger}
\affil{Palomar Observatory, California Institute of Technology, Mail
Stop 105-24, Pasadena, CA 91125}
\affil{e-mail: ccs,kla@astro.caltech.edu}
\author{\sc Mark E. Dickinson\altaffilmark{5}}
\affil{Department of Physics and Astronomy, The Johns Hopkins University,
N. Charles St., Baltimore, MD 21218}
\affil{e-mail: med@stsci.edu}
\author{\sc Max Pettini}
\affil{Royal Greenwich Observatory, Madingley Road, Cambridge, CB3 OEZ, UK}
\affil{e-mail: pettini@ast.cam.ac.uk}
\author{\sc Melinda Kellogg}
\affil{Palomar Observatory, California Institute of Technology, Mail
Stop 105-24, Pasadena, CA 91125}
\affil{e-mail: mk@astro.caltech.edu}

\altaffiltext{1}{Based on observations obtained at the Palomar Observatory,   
at the Kitt Peak National Observatory and at the W. M. Keck Observatory, which
is operated jointly by the California Institute of Technology and the
University of California.}
\altaffiltext{2}{Hubble Fellow.}
\altaffiltext{3}{Alfred P. Sloan Foundation Fellow.}
\altaffiltext{4}{NSF Young Investigator.}
\altaffiltext{5}{Allan C. Davis Fellow, also with the Space Telescope Science
Institute.}

\begin{abstract}

We have measured the angular correlation function \wth\ for a sample of 871
Lyman-break galaxies (LBGs) in five fields at redshift $z\sim 3$.  Fitting the
power-law $A_w\theta^{-\beta}$ to a weighted average of \wth\ from the five
fields over the range $12\simlt\theta\simlt 330$ arcsec, we find $A_w\sim 2$
arcsec$^{\beta}$ and $\beta\sim 0.9$. The slope is, within the errors, the
same as for galaxy samples in the local and intermediate redshift universe,
and a slope $\beta=0.25$ or shallower is ruled out by the data at the 99.9\%
confidence level.  Because $N(z)$ of LBGs is well determined from 376
spectroscopic LBG redshifts, the real-space correlation function can be
accurately derived from the angular one through the Limber transform. The
inversion of \wth\ is rather insensitive to the still relatively large
uncertainties on $A_{\omega}$ and $\beta$, and the spatial correlation length
is much more tightly constrained than either of these parameters. We estimate 
$r_0=3.3$\er{0.6}{0.7} (2.1\er{0.5}{0.4}) $\hh$Mpc (comoving) for $q_0=0.1$
(0.5) at the median redshift of the survey, $\bar z=3.04$ ($h$ is in units of
100 km s$^{-1}$ Mpc$^{-1}$ throughout this paper). The observed comoving
correlation length of LBGs at $z\sim 3$ is comparable to that of present-day
spiral galaxies and is only $\sim 50$\% smaller than that of present-day
ellipticals; it is as large or larger than any measured in recent
intermediate-redshift galaxy samples ($0.3 \simlt z \simlt 1$).  By comparing
the observed galaxy correlation length to that of the mass predicted from CDM
theory, we estimate a linear bias for LBGs of $b\sim 1.5$ ($4.5$) for 
$q_0=0.1$ ($0.5$), in broad agreement with our previous estimates based on
preliminary spectroscopy.  The strong clustering and the large bias of the
LBGs are consistent with biased galaxy formation theories and provide
additional evidence that these systems are associated with massive dark matter
halos.  The results of the clustering of LBGs at $z\sim 3$ emphasize that
apparent evolution in the clustering properties of galaxies may be due as much
to variations in effective light-to-mass bias parameter among different galaxy
samples as to evolution in the mass distribution through gravitational
instability.
\end{abstract}

\section{INTRODUCTION}

In most cosmological models, galaxies are expected to be biased tracers of the
underlying mass-density field, with the level of light-to-mass bias being a
function of the galaxy mass; more massive galaxies would tend to populate 
volumes of space with a higher overall mass density, and, as a consequence, 
would be characterized by a stronger spatial clustering than less massive 
systems (e.g. Kaiser 1984; Mo \& White 1996). Furthermore, the bias of galaxies
respect to the mass is expected to evolve with cosmic time as a result of 
gravitational growth of density perturbation and hierarchical merging 
(Matarrese \et 1997; Mann, Peacock \& Heavens 1997; Bagla 1997). Empirically,
it has been known for some time that different types of galaxies do indeed
cluster differently; numerous large galaxy redshift surveys (Davis \et 1988; 
Hamilton 1988; Santiago \& Da Costa 1990; Loveday \et 1995; Tucker \et 1996; 
Valotto \& Lambdas 1997) in the local universe have shown that early-type
galaxies (E/S0) are more strongly clustered than later types (Sp/Irr), with a
two-point correlation function that is generally steeper and a correlation
length $\sim 2$ times larger. A similar trend with the absolute luminosities
of the galaxies has also been observed (Park \et 1994).

In the past few years, several deep redshift surveys have probed field
galaxies in the intermediate-redshift universe (e.g. Lilly \et 1995; Cowie, Hu
\& Songaila 1995). These find similar clustering segregation as in the local
universe, with the redder and more luminous systems more strongly clustered
than their bluer and fainter counterparts (Le F\`evre \et 1996; Carlberg \et
1997), and detect apparent evolution in galaxy clustering, with a comoving
correlation length $r_0$ that is three times smaller at $z\sim 1$ than in
local samples.  If the galaxies in these samples at different redshifts all
had the same bias with respect to the mass distribution, then the observed
differences in galaxy clustering trace the evolution of mass clustering, and
could be used to constrain cosmology; however, it is difficult to understand 
the mix of galaxy masses included in magnitude limited surveys as a function 
of redshift. One might hope that a sample's bias would not depend strongly on
how it was selected, but if this were the case then different redshift surveys
would currently be in quantitative disagreement with each other (Carlberg \et
1997).  It seems likely, then, that a sample--dependent (both because of
redshift effects and selection criteria) light-to-mass bias could be at least
partly responsible for the observed ``evolution'' of galaxy clustering with
redshift, and in this case it would be difficult to draw cosmological
conclusions from those surveys.

Recently, it has become possible to identify large numbers of galaxies in a
narrow redshift range using photometric techniques (e.g. Connolly \et 1995,
Steidel \et 1996a, Madau \et 1996).  In contrast to traditional
magnitude--limited surveys, which contain a wide range of galaxies over a large
interval of time, and likely different mixtures of galaxies at different
redshifts, a sample selected in this way provides a snapshot of the locations
of similar galaxies over a small span of time.  As a result, the observed
clustering is much easier to interpret.  An example of a photometric redshift
technique is the Lyman-break technique (Steidel \& Hamilton 1993, Steidel,
Pettini, \& Hamilton 1995, Steidel \et 1996a, Giavalisco \et 1996, Madau \et
1996) which selects the brightest star-forming (and relatively dust-free) 
galaxies at high redshift.  It is still unclear what the lower redshift 
counterparts of these Lyman-break galaxies would be, and so one cannot easily 
draw cosmological conclusions by comparing the clustering strength of 
Lyman-break galaxies to the clustering strength of a lower-redshift sample; 
but we can use the sample for the more modest goal of constraining theories of
galaxy formation.  In particular there is a great deal of fruitful work to be 
done comparing the properties of this well-defined high-redshift sample with 
the predictions of numerical and semi-analytic models (Baugh \et 1998, Jing \&
Suto 1998, Governato \et 1998).

In a previous paper (Steidel \et 1998, Paper 1) we described a large
concentration of LBGs in redshift space discovered in one of our survey
fields, and argued that such a concentration would not exist in standard CDM
cosmogonies unless LBGs were very biased tracers of mass.  In the present
paper, we present a complementary angular clustering analysis of the LBG
candidates in 5 of our survey fields, which can be used in conjunction with
the spectroscopic redshift distribution to estimate the spatial correlation
function of the Lyman-break population at $z \sim 3$. Again, we will find that
these galaxies are much more strongly clustered than the mass would be
according to models of hierarchical structure formation. Such a strong
clustering of forming galaxies is actually in agreement with predictions of 
models in which LBGs are associated with relatively rare and massive dark
matter halos (e.g., Baugh \et 1997, Mo \& Fukugita 1996, Jing \& Suto 1998).

\section{LYMAN-BREAK GALAXIES}

The Lyman-break technique uses color selection to identify high-redshift
galaxies through multi-band imaging across the 912 \AA\ Lyman-continuum
discontinuity. Details of the technique have been presented elsewhere
(e.g. Steidel \& Hamilton 1993; Steidel, Pettini \& Hamilton 1995; Madau \et 
1996) and here we only briefly review them.  At $z \simgt 2.5$ the Lyman
limit is redshifted far enough into the optical window to be observable in
broad-band ground-based photometry.  By placing filters on either side of the
redshifted Lyman limit one can find high-redshift objects by their strong
spectral breaks.  In our implementation of the technique we use a custom
photometric system, $U_n\, G\, {\cal R}$ (Steidel \& Hamilton 1993) optimized
for selecting LBGs with $z \sim 3$.  An initial selection region of the
$[G-{\cal R},\, U_n-G]$ plane was chosen based on the expected colors of
moderately unreddened star-forming galaxies computed using stellar population
synthesis codes (Bruzual \& Charlot 1996), and including the effects of the
opacity of interstellar gas and intervening absorption by H~I (Steidel \et
1995; see also Madau 1995).  Our selection criteria were subsequently verified
and refined after extensive spectroscopy with the Low Resolution Imaging
Spectrograph (Oke \et 1995) on the W.M. Keck telescope.

In this paper an object is considered a Lyman-break galaxy if its colors
satisfy
$$(U_n-G)\ge 1.0+(G-{\cal R}) ; \quad (U_n-G)\ge 1.6 ;\quad  (G-{\cal R})\le
1.2,\eqno(1)$$ 
with an additional requirement ${\cal R}<25.5$ imposed to produce a reasonably
complete sample that is suitable for spectroscopic follow-up. Magnitudes are
in the $AB$ scale (Oke \& Gunn, 1983). We have found
that at least 75\% of the objects meeting these criteria are indeed
high-redshift galaxies. Figure 1 shows the redshift distribution $N(z)$ for
all 376 spectroscopically identified galaxies selected with these criteria. 
The median redshift is $\bar z=3.04$ and the standard deviation is
$\sigma_z=0.27$; approximately 90\% of the galaxies have $2.6\simlt z\simlt
3.4$, and none have $z<2.2$.  About 5\% of the objects meeting these criteria
are stars; almost all of these are brighter than ${\cal R}\sim 24$.  The
remaining 20\% of objects meeting these criteria have not been identified
because of low S/N.  Our success in obtaining a redshift has no obvious
dependence on luminosity or color.

For the purpose of measuring angular clustering we have restricted our sample
to candidates from the larger and deeper fields of our survey, whose salient
features are summarized in Table 1.  While each of the fields will be treated
independently in the analysis below, the surface density of faint galaxies,
and the median colors for all detected objects in each field, are consistent
with one another after correction for Galactic reddening. Each of our images
typically has seeing in the range $0\secpoint8-1\secpoint3$, and reaches
$1\sigma$ surface brightness fluctuations in 1 arcsec$^2$ apertures of $\sim
29.1$, $29.2$, and $28.6$ magnitudes/arcsec$^2$ in the $U_n$, $G$, and ${\cal
R}$ bands, respectively.  However, for objects fainter than ${\cal R}\approx
25$ small differences in the depth of the $U_n$ images can influence the
surface density of the faintest LBG candidates because of the large dynamic
range required to flag objects with significant continuum breaks.  For this
reason, we caution that the surface density of LBG candidates can have a small
dependence on the quality of the images in a particular field.

We are in the process of obtaining spectra of LBG candidates in each of these
fields using the Keck telescopes; in the present work, we make use only of the
redshift distribution for our spectroscopic survey as a whole.  We defer to a
future paper an analysis of the clustering in each field using the full
redshift information (Adelberger \et 1998, in preparation).

Most of the fields chosen for our LBG survey are high latitude, low Galactic
reddening fields that have been the subject of other faint galaxy studies;
particularly relevant to our choice of fields was the existence of deep WFPC-2
images in the {\it HST} archive.

The largest field is 1415+527, which is centered on a deep {\it HST}+WFPC-2
pointing and also contains several of the pointings of the ``Groth
strip''. The field has also been studied by Connolly \et (1997) for their
photometric-redshift technique, and by the Canada-France redshift survey
(Lilly \et 1995).  Our images were obtained using the prime focus camera on
the Mayall 4-m telescope at Kitt Peak during 1996 May and cover an area of
$15\times 15$ arcmin$^2$; the ${\cal R}$ image was supplemented with a mosaic
of images (in order to cover the whole KPNO 4m field of view) obtained at the
Palomar 5m Hale telescope with the COSMIC prime focus camera in 1997 March.

The 2237+116 field (DSF2237) was chosen by us as region with low Galactic
extinction and few bright stars which could be observed efficiently from both
Palomar and Keck observatories in the late summer/early fall observing season.
To our knowledge, no other faint galaxy studies have been conducted here. The
imaging data were obtained in 1997 August with the Palomar 5m telescope; the 
total region studied consists of two abutting pointings, each 9\arcm\ by 
9\arcm, aligned in the E-W direction.

Like the 2237+116 field, the 2215+000 field consists of two COSMIC abutting
pointings which have in this case been aligned along the North-South
direction. The images were obtained in 1995 August, 1996 August, and 1997
August.  The northern pointing is centered on the ``SSA22'' field of Cowie \et
(1995) and overlaps somewhat with the region studied as part of the CFRS
redshift survey ``22 hour'' field (Lilly \et 1995).  It also includes
several moderately deep {\it HST}+WFPC-2 pointings (Cowie \et 1996; Schade
\et 1995). In Paper 1 we have presented a preliminary analysis of LBG
spectroscopy in this region.

The 1234+625 (Hubble Deep Field, or HDF; Williams \et 1996) and 0050+123
(``Caltech Deep Field'', or CDF) fields were obtained as single pointings with
COSMIC.  The observations were carried out during March and April 1996 in the
former and during October 1996 in the latter, respectively. The HDF has been
extensively followed-up with Keck spectroscopy by several groups (Cowie \et
1997; Cohen \et 1996), including observations of Lyman-break galaxies
identified with {\it HST} (Steidel \et 1996b; Lowenthal \et 1997). In the CDF,
Cohen \et (1996) have obtained redshifts, primarily in the $0.3 \le z \le 1$
range, as part of the ``Caltech Deep Redshift Survey''; this field was chosen
by both Cohen \et and by us because of the existence of a deep WFPC-2 ``Medium
Deep Survey'' image at the center. For both the HDF and the CDF our images
cover a region much larger than, but including, the {\it HST} pointings.

\section {THE MEASURE OF \wth}

The angular correlation function \wth\ is defined in terms of the excess
probability over the random (Poisson) distribution of finding a companion in 
an angular shell of size $d\Omega$ placed at an angular separation $\theta$ 
from a selected galaxy, given the surface density of sources ${\cal N}$ 
(Peebles 1980):
$$dP = {\cal N}\, \Bigl[1+w(\theta)\Bigr]\, d\Omega.\eqno(2)$$

Usually \wth\ is measured by comparing the observed number of galaxy pairs at
a given separation $\theta$ to the number of pairs of galaxies independently 
and uniformly distributed over the same geometry as the observed field.  A 
number of statistical estimators of \wth\ have been proposed (e.g. see Landy 
\& Szalay 1993) in an attempt to minimize random and systematic errors.

We considered the two estimators
$$w(\theta) = {DD(\theta)\over DR(\theta)}-1\eqno(3)$$
and
$$w(\theta) = {DD(\theta)-2DR(\theta)+RR(\theta)\over RR(\theta)},\eqno(4)$$
proposed by Peebles (1980) and Landy \& Szalay (1993), respectively (PB and LS
in the following), where $DD(\theta)$ is the number of pairs of observed
galaxies with angular separations in the range $(\theta,\theta+\delta\theta)$,
$RR(\theta)$ is the analogous quantity for the homogeneous (random) catalog,
and $DR(\theta)$ is the number of observed-random cross pairs. Both of these
statistics produce estimates of \wth\ which are biased low by a factor (the
``integral constraint'') $I \simeq 1+O((\theta_0/\theta_{\rm max})^{\beta})$,
where $w(\theta_0)=1$ (Peebles 1974), but (as we will see) $\theta_0 \ll
\theta_{\rm max}$ and we will neglect this small correction. The properties of
the two statistics are discussed by Landy \& Szalay (1993). Relevant to our
analysis is the fact that the variance of the LS estimator is smaller than the
PB one, and is close to the variance of a Poisson distribution. In addition,
the LS estimator should be less sensitive to edge effects and the presence of
spurious variations of galaxy surface density (see below). We measured \wth\
using both estimators to test for the presence of such systematics.

We measured the correlation function using different techniques, as detailed
below. The analysis was carried out using two independently written programs,
finding virtually identical results.  In each case we computed \wth\ from each
individual field and then computed a weighted average using inverse variance
weighting; it made little difference if we used Poisson or bootstrap variance
(Ling, Barrow \& Frenk 1986). The error bars on the average correlation
function are the rms field-to-field variation in the estimated individual
\wth\ divided by $\sqrt{N_{\rm fields}}$; Poisson and bootstrap errors have
comparable size. Masking the regions around bright stars and galaxies where we
could have not detected LBGs had a negligible effect on the results.

We subsequently fitted the weighted average to the power law $A_\omega\theta^
{-\beta}$ with Levenberg-Marquardt nonlinear least-squares (Press \et
1992). To estimate confidence intervals on the parameters $A_{\omega}$ and
$\beta$, we generated a large ensemble of random realizations (100,000) of the
measured \wth, assuming normal errors, and calculated best fit parameter
values for each of these synthetic data sets (e.g. Press \et 1992 \S 15.6). We
found that the fitted parameters depend somewhat on the choice of the binning
used to compute \wth, and to take this additional source of uncertainties into
account we have included the effects of a randomly variable binning into the
Monte Carlo simulations. Table 2 lists the results. Because the fitted
parameters are strongly covariant, the 68\% confidence intervals are
misleadingly large, particularly the one relative to $A_{\omega}$; as we shall
see, the comoving correlation length $r_0$ is much more tightly constrained
than either $A$ or $\beta$ individually.

Figure 2 shows the weighted average estimates of \wth\ obtained from both the
PB and LS estimators. For clarity, the error bars are plotted separately from
the data points, in the upper part of the figure. A number of potential
sources of systematic errors can affect the estimate of \wth\ and we now
describe the techniques that we have adopted to test for their presence.

\subsection{Spurious Sources}

Systematic contamination from spurious sources, i.e. sources physically
unrelated to the Lyman-break galaxies, is relatively easy to take into
account.  Only $\sim 75\%$ of the spectroscopically observed Lyman-break
candidates have been shown to be at $z\sim 3$.  About 5\% are stars and the
remaining 20\% have not been identified.  The unidentified 20\% have spectra
and colors consistent with their being at $z\sim 3$, but in the worst case a
fraction $f\sim 0.25$ of the photometrically selected objects could lie at
redshifts outside of our primary selection window.  If these objects were not
clustered our estimate of \wth\ would be low by a factor of $1/(1-f)^2 \simeq
1.56$.  We will refer to this as the case of ``maximum contamination''.

\subsection{Field-to-Field Variations}

Particularly insidious are slight variations in detection probability across
the chip due to structure in flatfields, optical aberrations, software
performance, and so on.  If uncorrected, the resulting density gradients can
mimic galaxy clustering. We have tested for the presence of such effects by
using the locations of non-LBG galaxies instead of a uniform (random)
distribution when estimating $DR$ and $RR$. Because these faint galaxies are
intrinsically very weakly clustered, more than $10\times$ less clustered than
the LBGs at ${\cal R}\sim 25.5$ (e.g. Brainerd, Smail, \& Mould 1995), they
provide a reasonable approximation to a random distribution, and they have the
advantage of being subject to similar angular variations in detection
probability as our LBG sample. As can be seen in Table 2, the difference
between measures of \wth\ (from the same estimator) obtained using the random
distribution or the cross-correlation technique described above is comparable
to the random error and does not show any systematic trend, suggesting that 
the effect discussed here is not important at the current level of precision
of the data. 

Variations of sensitivity or in the photometric calibration (particularly in
the $U_n$ band) across individual fields are another possible source of
spurious clustering signal. This is because the density of objects with
$U_n-G$ and $G-{\cal R}$ colors close to edge of the color selection
``window'' is relatively high, and small variations in the photometry can
result in a significant number of galaxies being excluded from or included in
the sample.  Small spatial variations in sensitivity or color are most likely
to have the same dependence as the ${\cal R}$ detection probability, as they
would be due to small amounts of vignetting and/or variation in image quality
that would affect all bandpasses in a similar fashion, to first order.  The
quality of the CCDs used to obtain the images is high enough that spatial
variation in quantum efficiency, even in the $U_n$ band, are small enough so
as to be negligible in this context. Moreover, our photometric selection
criteria are conservative enough that any objects scattering into or out of
our selection window would also be bona fide LBGs, and their redshift
distribution would not be different enough to have a significant effect on the
clustering properties. That is, small spatial variations in the color zero
point could have only a second--order effect on the clustering properties,
given that the transformation of the angular into the spatial correlation 
function depends relatively weakly on the redshift distribution.

\subsection{Field Geometry}

Two of our fields, SSA22 and DSF2237, were each constructed from two abutting
CCD frames aligned along one direction in order to produce composite samples
covering larger area than that of the individual pointings. In general,
whether it is advantageous to estimate \wth\ from one large field with $N$
galaxies or from $M$ subfields with $N/M$ galaxies and then average the
results depends on the noise characteristics of the adopted estimator (see
Landy \& Szalay 1993 for a discussion). If the noise scales as $1/N$, like in
the case of the PB estimator, then it would make no difference. Analyzing the
abutting fields separately instead of together increases the total noise by a
factor of $M$ in the case of the LS estimator, because its variance scales as
$1/N^2$.  Furthermore, in both cases an additional error is also introduced by
the $1/M$ loss in the total pairs.  Thus, the combination of composite fields
and the LS statistics offers the possibility of extracting \wth\ from the data
maximizing the S/N, which is useful in cases like ours where the samples are
still relatively small and the clustering signal rather weak.

Unfortunately, fluctuations of the apparent galaxy surface density from field
to field due to images of differing depths, galactic extinction and reddening,
and so on, can introduce an artificial clustering signal in the observed \wth\
and cancel the benefits of using composite fields. Although a relatively large
overlapping region between the two individual images that compose these fields
allowed us to check the consistency of the photometry and selection criteria
between them, the possibility that the two sub-fields have slightly different
depth is difficult to test. Therefore, we have studied both cases of split and
composite fields with both the PB and LS statistics, each time using both the
random and cross-correlation techniques. The results are listed in Table 2. It
can be seen that, in every case, \wth\ in the case of composite fields is
shallower and with a larger correlation amplitude than that from the split
fields, whose parameters also have larger random errors. Although the
difference is comparable to the 1-$\sigma$ error bars, making the results
still consistent with each other, there is a possible presence of a systematic
error from this effect.

\subsection{The PB estimator vs the LS estimator}

Table 2 and Figure 2 show that the measures of \wth\ obtained using the PB
estimator are systematically larger than those from the LS one, with the
fractional difference between the two statistics being the largest at
separations of the order of $\sim 100$ arcsec. As a result, the correlation
function obtained using the PB statistics has a shallower slope and larger
correlation amplitude, and the random errors on the fitted parameters are
smaller. Again, such differences are comparable to the 1-$\sigma$ error bars
and, formally, the parameters derived from the two estimators agree with each
other. However, it is clear that the difference is systematic.  It is beyond
the scope of this paper to analyze the relative merits of the two estimators,
and here we simply report all the results, cautioning that the discrepancy
that we found, although comparable to the random errors, may imply that the 
two statistics are sensitive to the presence of low--level systematics in
different way.  

\section{THE INVERSION OF THE ANGULAR FUNCTION}

The angular correlation function \wth\ can be obtained from the spatial one,
$\xi(r)$, through the Limber transform if the galaxies' redshift distribution
$dN/dz$ is known (Peebles 1980; Efstathiou \et 1991). If the spatial function
can be modeled as
$$\xi(r)=(r/r_0)^{-\gamma}\times f(z),\eqno(5)$$
where $f(z)$ describes its redshift 
dependence, the angular function has the form 
$w(\theta)=A_w\theta^{-\beta}$, where $\beta=\gamma-1$ and 
$$A_w = C\, r_0^{\gamma}\, \int_{z_i}^{z_f} f(z)\, D_{\theta}^{1-\gamma}(z)\, 
\Bigg({dN\over dz}\Biggr)^2\, g^{-1}(z)\, dz\, \times 
\Biggl[\int_{z_i}^{z_f}\Biggl({dN\over dz}\Biggr)\, dx\Biggr]^{-2}\eqno(6)$$
(Efstathiou \et 1991). Here $D_{\theta}(z)$ is the angular diameter 
distance, 
$$g(z) = {c\over H_0}\bigl[(1+z)^2(1+\Omega_0z)^{1/2}\bigr]^{-1},\eqno(7)$$
and $C$ is a numerical factor given by
$$C = \sqrt{\pi}\, {\Gamma[(\gamma-1)/2]\over \Gamma(\gamma/2)}.\eqno(8)$$

The Lyman-break galaxies' redshift distribution is considerably narrower than
those of traditional flux-limited redshift surveys and spans a redshift
interval of only $\Delta z\sim 0.8$ (the corresponding cosmic-time interval is
$\approx 0.35\hh$ ($0.26\hh$) Gyr for $q_0=0.1$ ($q_0=0.5$). In this redshift
range and at the spatial scales considered here (larger than a few Mpc), the
evolution of cosmic structure is largely driven by the growing mode of linear
perturbations $D(t)$. The front-to-end fractional variation of $D(t)$ in the
above redshift range is $\sim 13$ (18\%), which is an upper limit to the
effective variation of the correlation length in our sample because of the
peaked redshift distribution. It is reasonable, therefore, to expect little
evolution of LBG clustering in our sample over so short a time.  In this case
the function $f(z)$ above can be taken out of the integral and the quantity
$r_0(z)=r_0\, f(z)$ is then the correlation length at the epoch of
observations.

We list in Table 3 the comoving correlation lengths $r_0(z)$ at $\bar z=3.04$
obtained through Eqn. (6) using the average \wth\ and the redshift
distribution $N(z)$ plotted in Figure 1. The 68\% confidence intervals were
computed using Monte Carlo simulations.  As mentioned above, the correlation
length turns out to be much more tightly constrained than the individual
parameters of the angular correlation function, with a typical fractional
error of $\approx 20$--30\% at the 1-$\sigma$ level. This is enough precision
to show the effects of the systematic differences between the values of $r_0$
obtained from the PB and LS statistics and between keeping the two adjacent
pointings in the SSA22 and DSF2237 fields as independent or joining them into
two composite fields, respectively. Figures 3a and 3b shows the distribution
of values of $r_0$ obtained combining together the measures of $r_0$ from the
Monte Carlo simulations for the PB (thin continuous histogram) and LS
estimators (broken histogram) respectively, for the two values of $q_0$
considered in the paper. We defer a deeper analysis of the systematics to
future papers. At this time, because the differences between all the various
measures of $r_0$ that we have obtained are still comparable to the 1-$\sigma$
random errors, we list them all. As our fiducial measure and error bar we
adopt the median of the distribution obtained by merging the PB and LS Monte 
Carlo distributions and its corresponding 68\% confidence interval. These are
$r_0=3.3$\er{0.6}{0.7} and $r_0=2.1$\er{0.5}{0.4} $\hh$ Mpc for $q_0=0.1$ and
$q_0=0.5$, respectively, and the histograms of these distributions are plotted
in Figure 3 as thick continuous lines. We have used these values to produce
the plots in Figure 4. Our fiducial value of $\beta$, obtained in a similar
fashion is $\beta=0.98$\er{0.28}{0.32} and the histograms of the Monte Carlo
simulations for the PB, LS and combined samples are plotted in Figure
3c. Finally, we remind that all values of correlation length would be
approximately 25\% higher in the case of ``maximum contamination.''

We can estimate the bias of these galaxies by comparing their correlation
function $\xi_g$ to the correlation function of the mass $\xi_m$:
$$b(r)=\sqrt{{\xi_g(r)\over \xi_m(r)}}.\eqno(9)$$ 
Although (as eq. 9 shows) the bias is in principle a function of scale, our 
constraint on the power-law exponent $\gamma\equiv 1-\beta$ is relatively weak
(see Table 3), and we can only estimate a ``typical'' value of the bias over 
the scales of a few Mpc which are probed here. In practice, we use the ratio 
of the correlation length of the LBGs to that of $\xi_m(r)$ predicted by the 
CDM theory to compute the bias, which is therefore relative to $r=1$ $\hh$Mpc. 
Using a CDM power-spectrum with shape parameter $\Gamma^* = 0.25$, claimed to 
fit the shape of the local large-scale structure very well (Peacock 1997), and
normalization of Eke, Cole, \& Frenk (1996), we estimate $b\sim 1.5$ $(4.5)$ 
for $q_0=0.1$ $(0.5)$. Choosing $\Gamma^*=0.20$ results in $b\sim 1.5$ $(5)$, 
while adopting the normalization of White, Efstathiou \& Frenk (1993) results 
in $b\sim 1$ $(4)$. 

Overall, these bias values are slightly lower than those estimated in Paper 1,
but are consistent with them at the $\sim$ 10\% confidence level.

\section {DISCUSSION}

The high efficiency of the Lyman-break technique and the relatively narrow
range of redshifts/cosmic time that it probes make angular clustering a
particularly economic means to study large-scale structure at high redshifts, 
once the redshift distribution of the galaxy candidates has been measured. Not
only is one free from securing a complete spectroscopic follow-up of the
candidates, but the systematics due to selection effects are easier to handle
than than those that affect studies of spatial clustering using the full
redshift information. Our two main conclusions from an analysis of the angular
clustering of Lyman-break galaxies in the redshift range $2.5 \simlt z \simlt
3.4$ are that these systems are strongly clustered and that their correlation
function has a slope which is as steep or steeper than that of local galaxies.

In this section, we discuss the implications of these conclusions in turn. 

\subsection{The Slope of The Correlation Function}

The correlation function of the Lyman-break galaxies has a slope that is
comparable to or steeper than that measured at intermediate and low
redshifts. Table 2 shows the values of $\beta$ obtained from the various
techniques discussed above for each of the PB and LS estimators, respectively.
Combining all the PB Monte Carlo distributions together, we found $\beta_{PB}=
0.80$\er{0.19}{0.25}, where the error bars are the 68\% confidence interval.
Within the errors, this is the same value found for field galaxies in flux
selected surveys. The LS estimator returns a slightly steeper correlation
function, with $\beta_{LS}=1.14$\er{0.23}{0.29}, comparable to that of the
earliest (E/S0) and/or most luminous local galaxies (e.g. Loveday \et
1995). Combining all the Monte Carlo distributions together we found
$\beta=0.98$\er{0.28}{0.32}. The distribution of the PB slope rules out
$\beta=0.25$ or shallower at the 99.9\% confidence level, while the
distribution of the LS slopes rules out $\beta=0.49$ or shallower at the same
confidence level.

The evolution of the slope of the correlation function of the mass (or,
equivalently, that of the power spectrum at small scales) has a pronounced
dependence on $\Omega$. For a CDM-like power spectrum, it depends very weakly
on the shape parameter $\Gamma^*$ and, for flat models, on the normalization.
As Eqn.(9) shows, the slope of $\xi_g(r)$ differs from that of $\xi_m(r)$
because of the dependence of the bias parameter $b(r)$ with the spatial scale.
The form of $b(r)$, its dependence on galaxy properties and how it evolves
with redshift are still subjects of discussion (e.g. Mann, Peacock \& Heavens
1997; Bagla 1997). If the scale dependence of $b(r)$ for the LBGs over the
spatial scales probed by our correlation analysis, namely $1\simlt r\simlt 10$
$\hh$Mpc, is similar to that of the local galaxies, then our measures of
$\gamma=\beta+1$ are inconsistent with $\xi_m(r)$ from the CDM theory if
$\Omega=1$. With our choice of $\Gamma^*=0.25$ we found $\gamma_m=1.25$ (over
the range $1<r<10$ $\hh$Mpc), independently of the normalization. As mentioned
above, the dependence on $\Gamma^*$ is very weak. If $\Gamma^*=0.1$ then
$\gamma_m=0.98$, while if $\Gamma^*=0.6$, then $\gamma_m=1.14$. Using the
slope measured from the PB estimator, the steepest CDM slope ($\gamma_m=1.25$)
is ruled out by our data at the 99.90\% confidence level. Using the measures
from the LS estimator, it is ruled out at the 99.991\% confidence level. Open
CDM models with the same parameters as above produce slopes in the range
$1.6<\gamma_m<2.1$ (in open models the slope of $\xi_m(r)$ has a more
pronounced dependence on the normalization), which are all consistent with our
data.

The above computations assume a bias constant with spatial scale. We stress,
however, that the evolution of the slope of the correlation function is useful
for constraining cosmological models only if the dependence of the bias with
the spatial scale and its evolution with redshift are known.  The function 
$b(r)$ also depends on the properties of the halos, which further complicates
the interpretation of the data because of the difficulty of establishing an
evolutionary sequence between the systems observed at high redshifts and the
local galaxies. Bagla's (1997) N-body simulations seem to suggest that the
bias will not be strongly scale-dependent  ---his $b(r)$ for $M\ge 2\times
10^{12}$ \msun\ halos at $z=0$ in standard CDM has a power-law slope of only
$\sim-0.18$--- and if $b(r)$ for Lyman-break galaxies is similarly flat, our
conclusions about the slope would not be importantly changed. But until more
is known about the scale-dependence of the bias they will remain speculative. 

\subsection{Spatial Clustering and its ``Evolution''}

The LBGs at $z\sim 3$ are characterized by strong spatial clustering, with a
co-moving correlation length of $\sim 3-4h^{-1}$ Mpc for a low matter density
($\Omega_0=0.2$) Universe. The correlation length would be $\sim 25$\% higher
if the ``maximum contamination'' applies. This is comparable to the clustering
of present-day spiral and IRAS galaxies, and a factor of $\approx 2$ smaller
than that of present-day ellipticals.

A simple comparison of the observed clustering properties with the expected
clustering of a suitably normalized CDM density field, as shown in Figure 3b,
suggests that, in the context of such models, the LBGs must be substantially
biased with respect to the dark matter, with higher bias required in models
with higher matter density. This result is in qualitative agreement with our
conclusions in Paper 1 on the basis of the clustering in redshift space in one
of our survey fields, although the redshift-space analysis may imply somewhat
higher bias (on slightly different scales) than the correlation function
analysis presented here. As discussed there, the strong clustering and the
large bias of the Lyman-break galaxies are consistent with biased galaxy 
formation theories and provide additional evidence that these systems are
associated with massive dark matter halos.

One might be tempted to try to fit the Lyman break galaxy clustering
properties onto an general evolutionary sequence for galaxy clustering versus
cosmic epoch.  Figure 4a shows the galaxy clustering strength measured from
various redshift surveys, including the LBGs, as a function of redshift. To
quantify the clustering strength we have used the function $r_0^{\gamma}\times
(1+z)^3$, where now $r_0$ is in {\it proper} coordinates. Figure 4b shows the
corresponding plot of the bias derived using the CDM mass correlation
function, which we have computed using the non-linear code for the evolution
of the power spectrum by Peacock (1997).  We used the shape parameter
$\Gamma^*=0.25$ and normalized the power spectrum to $\sigma_8=1.0$ for the
open model and $\sigma_8=0.5$ in the Einstein-de Sitter case. The error bars
have been computed by propagating the errors in the measure of the correlation
length, but it is clear that the dominant uncertainty is in the choice of
normalization of the theoretical curve.

As Figure 4 suggests, the variations in the value of the ``effective'' bias
from sample to sample complicate the interpretation of the apparent evolution
of galaxy clustering as due to the gravitational growth of structures,
preventing us from deriving information on the clustering evolution of the
mass. For example, the data show that the traditional power-law model
$\xi(r,z)=\xi_0(r)\times (1+z)^{-(3+\epsilon)}$ used to describe the
gravitational evolution of clustering (Peebles 1980) is not a good
representation for any value of $\epsilon$ over the redshift interval $0\simlt
z\simlt 3$. In fact, it is clear from N-body simulations (e.g., Brainerd \&
Villumsen 1993, Bagla 1997) that the behavior of the clustering of {\it
halos}, as opposed to that of the overall mass, will have a non-trivial
dependence on redshift that is not accounted for in the ``$\epsilon$''
models. When one also notes the uncertainties in the way in which dark halos
are effectively sampled in any given redshift survey, we question the ultimate
usefulness of fitting values of this parameter to the results of redshift
surveys over substantial redshift baselines.

Lyman-break galaxies at $z \sim 3$ represent both a large jump in redshift and
a substantially different detection/selection technique than has been used
previously in most of galaxy clustering analyses. It is highly likely that a
similar selection criterion applied to nearby galaxies would result in a very
different correlation function than for (nearby) optically-selected galaxies.
If star formation progresses to less massive systems with time (e.g., Cowie
\et 1997, Giavalisco \et 1996), then a Lyman-break galaxy sample over a large
range of redshifts would likely exhibit a gradually {\it diminishing}
clustering strength with time, rather than a monotonically increasing
clustering strength due to growth of the overall mass fluctuations due to
gravitational instability. Our results emphasize that the interpretation of
apparent ``evolution'' of the clustering properties of galaxies in deep
surveys (e.g., Efstathiou 1995; Brainerd \et 1995, etc.) cannot be directly
interpreted as a reliable measure of the overall growth of structure. In a
typical survey limited by apparent magnitude, the mix of galaxy types and the
relative sensitivity of the survey to stellar mass, star formation rate, and
bolometric luminosity (and the complicated manner in which these quantities
are related to overall galaxy mass) will be strongly redshift-dependent.  It
is apparent now, if it has not always been, that the clustering properties of
galaxies at moderate to high redshift cannot be used as a cosmological tool
unless one is prepared to simultaneously understand where galaxies form and
how they evolve relative to the underlying distribution of dark matter
---cosmology and galaxy formation cannot be understood independently in this
context.

These problems argue strongly in favor of a focused approach to studies of
large-scale structure over substantial redshift baselines, and call into
question the very use of the phrase ``evolution of galaxy clustering''.  The
results found in this paper reiterate that selecting a particular class of
objects as defined by their observational properties, over a relatively narrow
range of cosmic time, may offer the best hope of constructing meaningful
samples for clustering analyses that can be compared with theoretical
predictions (e.g. Le Fevre \et 1996; Carlberg \et 1997).  At very high 
redshifts the specific mechanisms of galaxy formation are expected to affect
the clustering of forming systems the most. In the case of the $z \sim 3$
LBGs, the interval of cosmic time spanned by such samples are small enough,
and the selection criterion uniform enough, that even if it may be quite
model-dependent to ``map'' them onto samples of galaxies selected differently
at other redshifts, one might at least be confident of measuring something
specific that can be easily compared to the predictions of models or 
simulations (e.g. Adelberger \et 1998, Giavalisco \et 1998b, in prep.).

Because the LBG selection technique (or other equivalent to
it) is sensitive to those galaxies that are the most actively star-forming
systems (at whatever redshift/epoch probed), understanding the evolution of
the clustering with redshift would involve not only the modeling of the
underlying dark matter distribution, which is now relatively straight-forward
using N-body simulations, but also a detailed understanding of which types of
objects (i.e., which ``halos'') would be harboring star formation at
detectable levels as a function of time.  Simulations that include star
formation (e.g., Baugh \et 1997, Weinberg \et 1997) can make direct
predictions of the clustering properties of objects subject to a star
formation rate threshold, based on semi-analytic treatment of star formation
within dark matter halos.  The observed clustering strength of LBGs is close
to predictions of these models (particularly after possible corrections for a
small amount of contamination in the photometrically selected LBG sample), and
similar numbers result naturally from pure N-body or analytic models that
assume a mapping of the most massive virialized halos at $z\sim 3$ to objects
likely to be visible because of their star formation (Bagla 1997, Mo \&
Fukugita 1996, Jing \& Suto 1997).

Our current measurements can be improved upon in a number of ways. First,
obtaining photometric identifications of LBGs over much larger fields would
result in much better independent constraints on the amplitude and slope of
the angular correlation function. Secondly, a full real-space analysis of the
clustering properties of the LBGs in the same fields will offer a much higher
signal-to-noise ratio (and a different dependency on cosmological parameters)
than can be attained from the angular distribution of candidates coupled with
an empirical redshift selection function.  The real-space analysis requires a
more careful treatment of observational selection effects, and reasonably
complete spectroscopy, and we have deferred this to a future paper (Adelberger
\et 1998).

\section{SUMMARY}

We have measured the angular correlation function \wth\ of Lyman-break
galaxies at redshift $z\sim 3$.  Fitting the power-law $A_w\theta^{-\beta}$ to
a weighted average of \wth\ from the five fields over the range
$12\simlt\theta\simlt 330$ arcsec, we find $A_w\sim 2$ arcsec$^{\beta}$ and
$\beta\sim 0.9$. The slope is, within the errors, the same as for galaxy
samples in the local and intermediate redshift universe, and a slope
$\beta=0.25$ or shallower is ruled out by the data at the 99.9\% confidence
level.

Because the redshift distribution $N(z)$ of LBGs is well determined from 376
spectroscopic redshifts, we have derived the real-space correlation function
from the angular one through the Limber transform. The inversion of the \wth\
is rather insensitive to the still relatively large uncertainties on
$A_{\omega}$ and $\beta$, and the spatial correlation length $r_0$ is much
more tightly constrained than these.  Using Monte Carlo simulations to derive
the $1\sigma$ error bars from the 68\% confidence interval, we estimate
$r_0=3.3$\er{0.6}{0.7} (2.1\er{0.5}{0.4}) $\hh$Mpc (comoving) for $q_0=0.1$
(0.5) at the median redshift of the survey, $\bar z=3.04$. Thus, the observed
comoving correlation length of LBGs at $z\sim 3$ is comparable to that of
present-day spiral galaxies and is only $\sim 50$\% smaller than that of
present-day ellipticals; it is as large or larger than any measured in recent
intermediate-redshift galaxy samples ($0.3 \simlt z \simlt 1$).

By comparing the observed correlation length of LBGs to that of the mass
predicted from CDM theory, we have estimated a linear bias for LBGs of $b\sim
1.5$ ($4.5$) for $q_0=0.1$ ($0.5$), in broad agreement with our previous
estimates based on preliminary spectroscopy (Paper 1). The strong clustering
and the large inferred bias of the LBGs are consistent with biased galaxy
formation theories and provide additional evidence that these systems are
associated with massive dark matter halos.  

The evolution of the slope of the correlation function of the mass (or,
equivalently, that of the power spectrum at small scales) has a pronounced
dependence on $\Omega$. If the biasing parameter is a weak function of the
spatial scale, the measured slope of the correlation function of LBGs,
$\gamma=1.98$\er{0.28}{0.32}, is inconsistent with the predictions of the
standard CDM theory with $\Omega=1$ at the 99.9\% confidence level. N-body
simulations seem to suggest that the bias will not be strongly 
scale-dependent; however, until more is known about the scale-dependence of
the bias, this conclusion will remain speculative.

The results of the clustering of LBGs at $z\sim 3$ emphasize that apparent
evolution in the clustering properties of galaxies may be due as much to
variations in effective light-to-mass bias parameter among different galaxy
samples as to evolution in the mass distribution through gravitational
instability. Our study shows that the traditional power-law model 
$\xi(r,z)=\xi_0(r)\times (1+z)^{-(3+\epsilon)}$ traditionally used to describe
the gravitational evolution of clustering (Peebles 1980) is not a good
representation for any value of $\epsilon$ over the redshift interval $0\simlt
z\simlt 3$.

\vskip1cm

We are grateful to an anonymous referee for useful comments and suggestion
that have improved our paper.  We would like to thank the staff at the
Palomar, Kitt Peak and Keck observatories for their invaluable help in
obtaining the data that made this work possible.  We are greatly indebted to
Ed Carder at KPNO for understanding the cause of the $CuSO_4$ leakage in the
$U_n$ filter and his superb work in repairing it.  We have benefited from
stimulating conversations with Ray Carlberg, John Peacock and Cedric Lacey,
who have also kindly made available to us their theoretical models of CDM
clustering evolution. We are grateful to Richard Ellis for his useful comments
on an early version of the paper. MG has been supported by the Hubble
Fellowship program through grant HF-01071.01-94A awarded by the Space
Telescope Science Institution, which is operated by the Association of
Universities for Research in Astronomy, Inc. under NASA contract NAS
5-26555. CCS acknowledges support from the U.S. National Science Foundation
through grant AST 94-57446, and from the Alfred P. Sloan Foundation.

\newpage
\begin{deluxetable}{llcrrr}
\tablewidth{0pc}
\scriptsize
\tablecaption{The Observed Fields}
\tablehead{
\colhead{\#} & 
\colhead{Field} & 
\colhead{Size\tablenotemark{a}} & 
\colhead{N\tablenotemark{b}} & 
\colhead{${\cal N}$\tablenotemark{c}} & 
\colhead{$\alpha$\tablenotemark{d}} }
\startdata
1 & 0050+123 (CDF)      &  $8.8\times  8.9$ &  80 & 1.02 & 59.0 \nl
2 & 1234+625 (HDF)      &  $8.6\times  8.7$ & 104 & 1.39 & 50.9 \nl
3 & 1415+527 (Westphal) & $15.1\times 15.1$ & 293 & 1.29 & 52.9 \nl
4 & 2215+000 (SSA22)    &  $8.6\times 17.6$ & 186 & 1.23 & 54.1 \nl
4a& 2215+000 (SSA22a)   &  $8.6\times  8.9$ &  87 & 1.14 & 56.3 \nl
4b& 2215+000 (SSA22b)   &  $8.6\times  9.0$ &  99 & 1.28 & 53.1 \nl
5 & 2237+114 (DSF2237)  & $17.4\times 10.1$ & 208 & 1.18 & 55.2 \nl
5a& 2237+114 (DSF2237a) & $ 9.2\times 10.1$ &  86 & 0.93 & 62.4 \nl
5b& 2237+114 (DSF2237b) & $ 9.0\times 10.1$ & 127 & 1.40 & 50.8 \nl
\enddata
\tablenotetext{a}{In units of arcmin$^2$.}
\tablenotetext{b}{Number of LBG candidates with ${\cal R}\le 25.5$.}
\tablenotetext{c}{Surface density at ${\cal R}\le 25.5$; galaxies per 
arcmin$^2$.}
\tablenotetext{d}{Mean intergalaxy angular separation at ${\cal R}\le 25.5$; 
arcsec.}
\end{deluxetable}
\newpage
\begin{deluxetable}{lcc}
\tablewidth{0pc}
\scriptsize
\tablecaption{The Fitted Parameters}
\tablehead{
\colhead{Estimator} & 
\colhead{$A_w$\tablenotemark{a}} & 
\colhead{$\beta$\tablenotemark{b}} }
\startdata
PB rand       & 1.3\er{0.7}{1.2} & 0.7\er{0.1}{0.1} \nl
PB rand-split & 2.1\er{1.2}{3.1} & 0.9\er{0.2}{0.2} \nl
PB xcor       & 1.0\er{0.6}{1.2} & 0.7\er{0.2}{0.2} \nl
PB xcor-split & 2.3\er{1.5}{4.7} & 1.0\er{0.2}{0.3} \nl
PB all        & 1.5\er{0.9}{2.3} & 0.8\er{0.2}{0.3} \nl
LS rand       & 3.4\er{2.1}{5.5} & 1.1\er{0.2}{0.3} \nl
LS rand-split & 4.1\er{2.8}{10.0}& 1.2\er{0.3}{0.3} \nl
LS xcor       & 3.8\er{2.2}{5.1} & 1.1\er{0.2}{0.2} \nl
LS xcor-split & 6.2\er{4.0}{12.3}& 1.3\er{0.2}{0.3} \nl
LS all        & 4.1\er{2.6}{7.8} & 1.1\er{0.2}{0.3} \nl
PB + LS all   & 2.5\er{1.7}{5.5} & 1.0\er{0.3}{0.3} \nl
\enddata
\tablenotetext{a}{Angular correlation amplitude in arcsec$^{\beta}$}
\tablenotetext{b}{Angular correlation slope}
\end{deluxetable}
\newpage
\newpage
\begin{deluxetable}{lcc}
\tablewidth{0pc}
\scriptsize
\tablecaption{The Correlation Length\tablenotemark{a}}
\tablehead{
\colhead{} &
\colhead{$r_0$} &
\colhead{$r_0$} }
\startdata
$q_0=0.1$~~~&       PB         &        LS        \nl
rand        & 3.9\er{0.4}{0.4} & 2.9\er{0.5}{0.4} \nl
rand-splt   & 3.5\er{0.5}{0.5} & 2.7\er{0.6}{0.5} \nl
xcor        & 3.7\er{0.6}{0.5} & 3.3\er{0.4}{0.5} \nl
xcor-splt   & 3.1\er{0.5}{0.5} & 2.8\er{0.5}{0.5} \nl
all         & 3.6\er{0.6}{0.5} & 2.9\er{0.6}{0.5} \nl
\hline
$q_0=0.5$~~~&                  &                  \nl
rand        & 2.5\er{0.3}{0.3} & 1.8\er{0.3}{0.3} \nl
rand-splt   & 2.3\er{0.4}{0.2} & 1.7\er{0.3}{0.3} \nl
xcor        & 2.4\er{0.4}{0.3} & 2.1\er{0.3}{0.3} \nl
xcor-splt   & 2.0\er{0.3}{0.3} & 1.8\er{0.3}{0.3} \nl
all         & 2.3\er{0.4}{0.3} & 1.9\er{0.4}{0.3} \nl
\enddata
\tablenotetext{a}{Comoving coordinates, in units of $\hh$ Mpc}
\end{deluxetable}
\newpage
\begin{figure}
\figurenum{1}
\epsscale{1.0}
\plotone{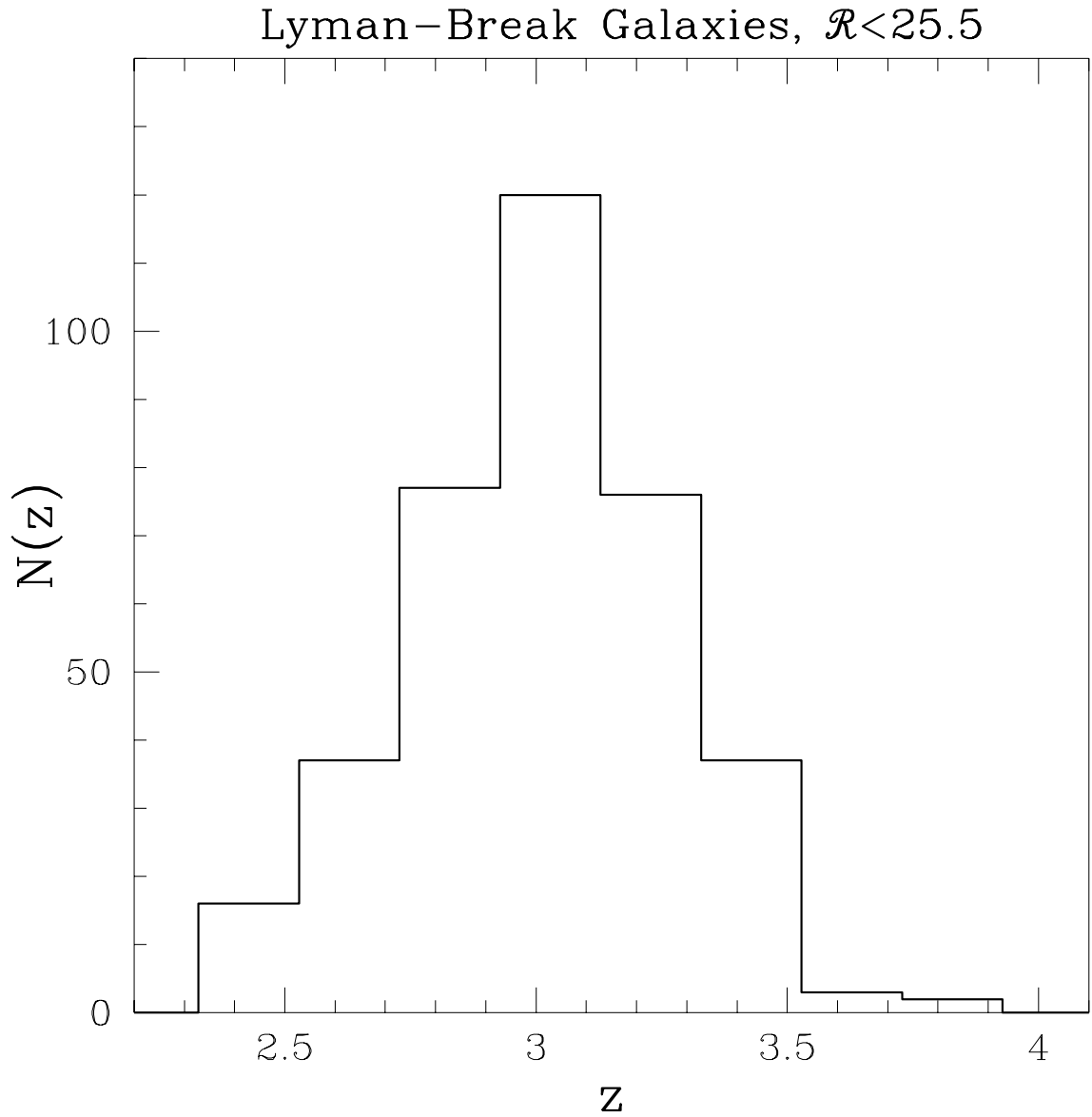}
\caption{The redshift distribution function $N(z)$ of the 376 LBGs used in the
correlation analysis. The bin size is $\Delta z=0.2$. The interval $2.6\simlt 
z\simlt 3.4$ contains $90$\% of the galaxies.}
\end{figure}
\newpage
\begin{figure}
\figurenum{2}
\epsscale{1.0}
\plotone{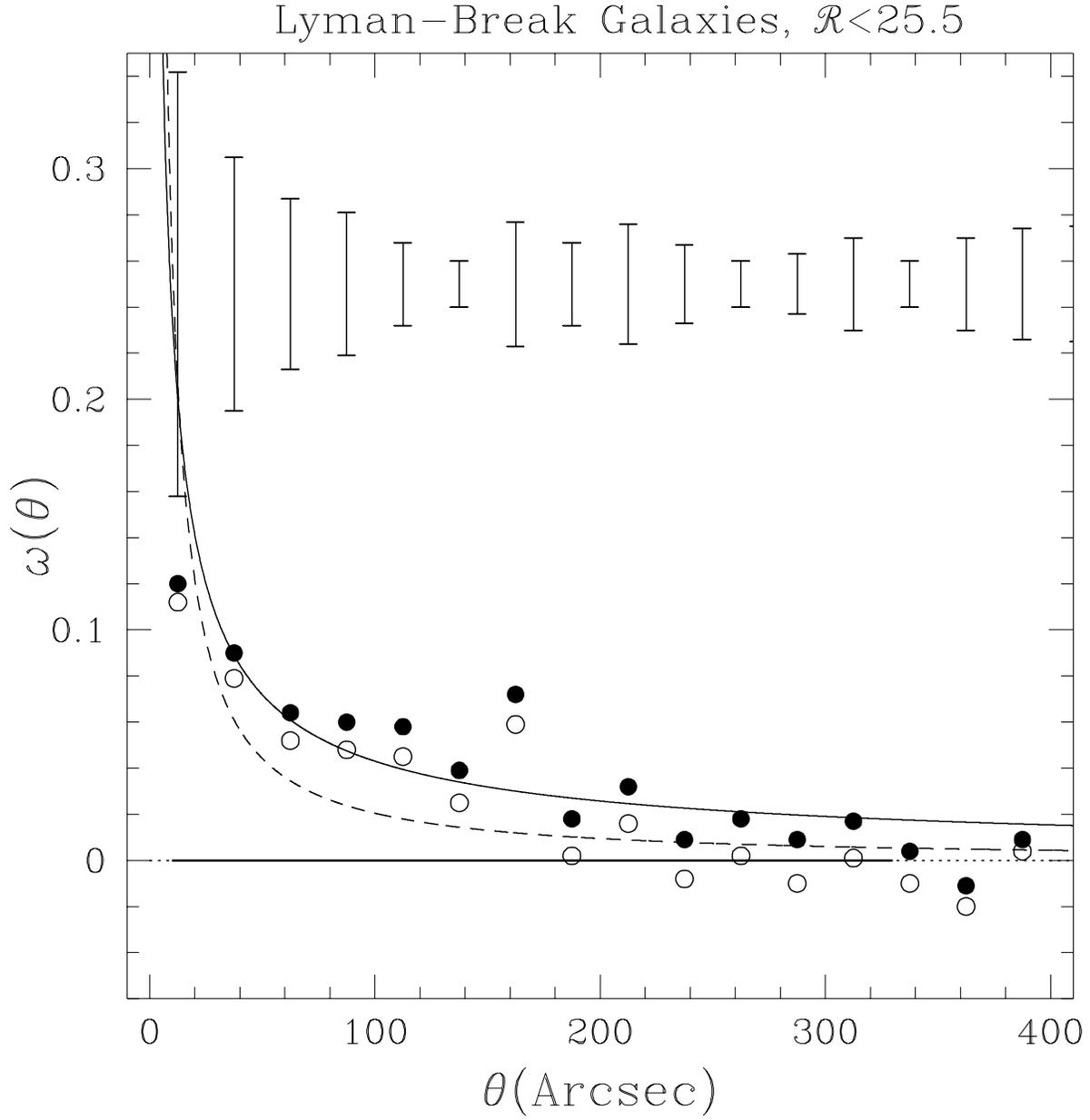}
\caption{Weighted average angular correlation function of LBGs. The filled
points are from the PB estimator, the open points from the LS one. The error
bars are shown on the top of the figure. The continuous line is the best-fit
power law to the PB data points, the dotted line is the fit to the LS. The 
thick horizontal continuous segment on the x axis marks the angular range over
which we computed the fits.} 
\end{figure}
\newpage
\begin{figure}
\figurenum{3}
\epsscale{1.0}
\plotone{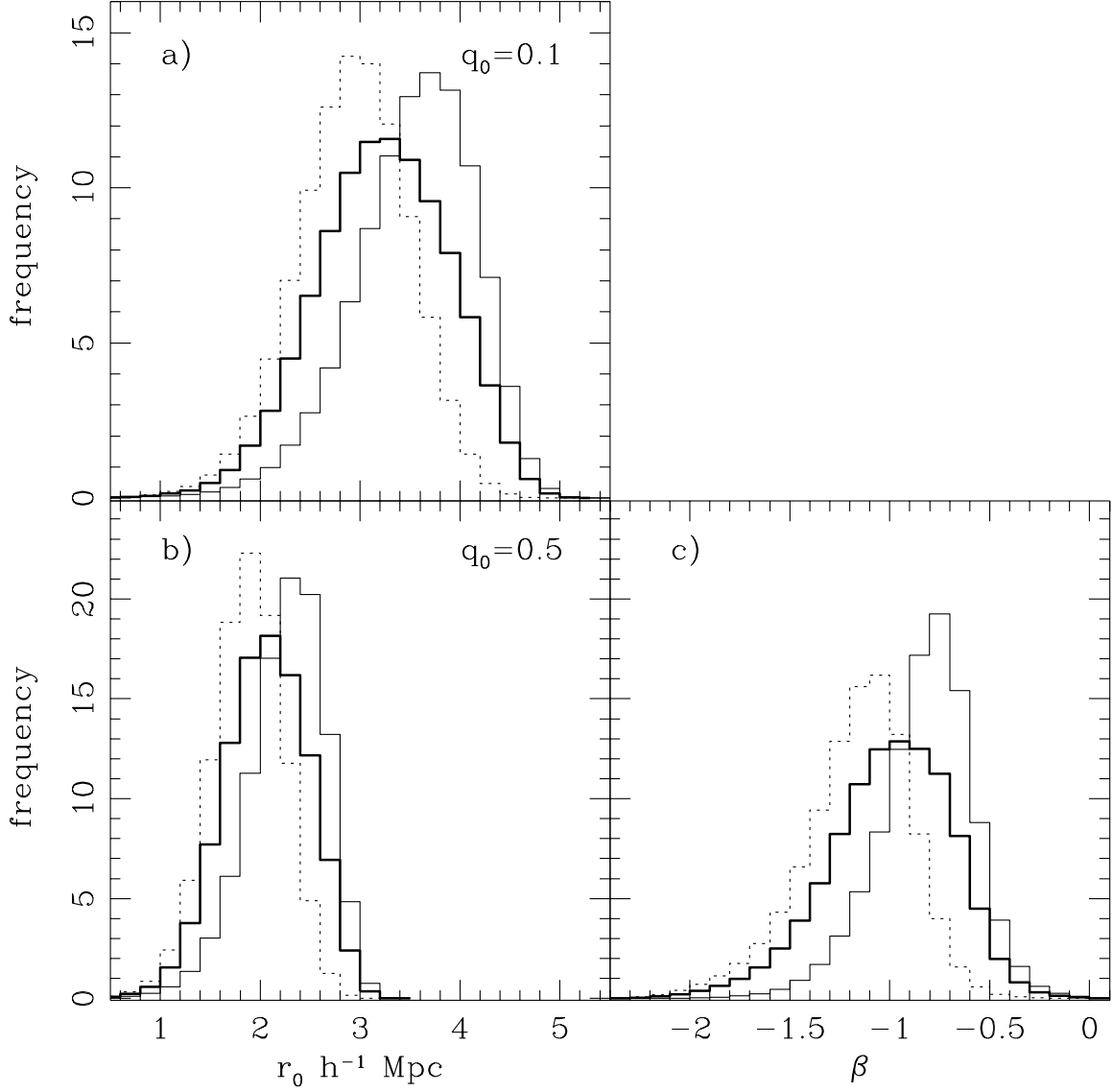}
\caption{The histogram of the correlation length $r_0$ ({\bf a)} and {\bf b)})
and of the slope $\beta$ ({\bf c)}) from the Monte Carlo simulations. The 
thin continuous line is for the PB estimator, the broken line is for the LS
estimator. For each estimator, the Monte Carlo distributions corresponding to
the four different measures listed in Table 3 have been merged together. The
thick continuous line is the distribution of all the PB and LS Monte Carlo
samples merged together.}  
\end{figure}
\newpage
\begin{figure}
\figurenum{4}
\epsscale{1.0}
\plotone{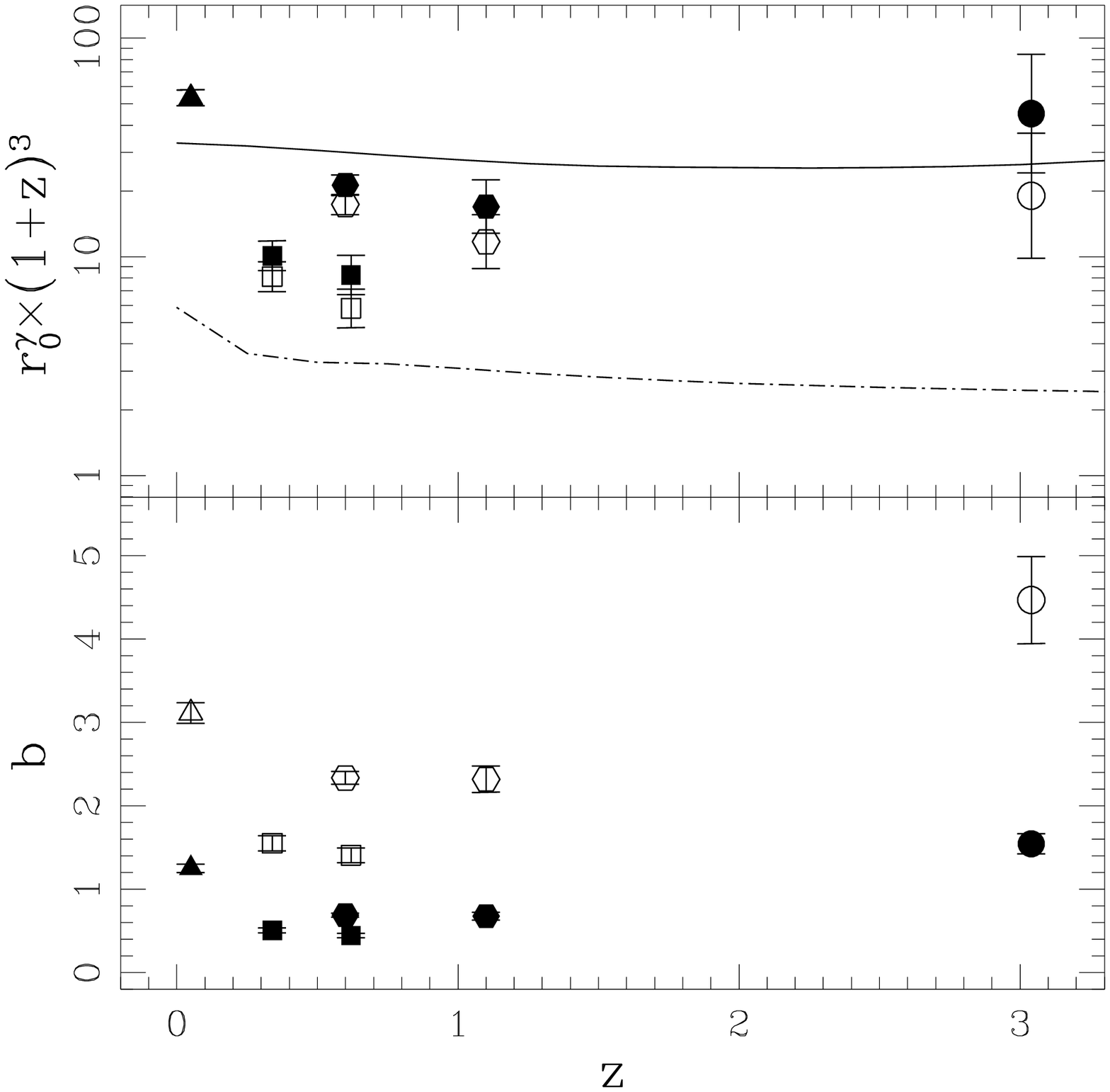}
\caption{{\bf a)} The strength of galaxy clustering as a function of 
redshift. Filled symbols are for $q_0=0.1$, open symbols for $q_0=0.5$. 
Triangles are APM data (Loveday \et 1995), squares are CFRS data (Le F\`evre 
\et 1996), hexagons Keck K-band data (Carlberg \et 1997), and circles are
the LBG data. The continuous solid and dashed lines are the expectations from
the CDM theory with $\Gamma^*=0.25$, $\sigma_8=1.0$ and  0.5 for the two case
of $q_0=0.1$ and 0.5, respectively. 
{\bf b)} Linear bias as a function of redshifts for the same data sets as a),
assuming the CDM correlation function. Symbols are defined as above.}
\end{figure}

\begin{references}

\reference{} Adelberger, K. L., Steidel, C. C., Giavalisco, M., Dickinson, M. 
             E., Pettini, M., \& Kellogg, M. 1998, ApJ, submitted.
\reference{} Bagla, J. S. 1997, MNRAS, submitted, astro-ph/9711081
\reference{} Baugh, C. M., Cole, S., Frenk, C. S., \& Lacey, C. G. 1998,
             ApJ, in press, astro-ph/9703111
\reference{} Brainerd, T. G. \& Villumsen, J. V. 1994, ApJ, 431, 477
\reference{} Brainerd, T. G., Smail, I., \& Mould, J. 1995, MNRAS, 275, 781
\reference{} Carlberg, R. G., Cowie, L. L., Songaila, A., Hu, E. M. 1997, 
             ApJ, 484, 538
\reference{} Cohen, J. G., Cowie, L. L., Hogg, D. W., Songaila, A., Blandford,
             R., Hu, E. M., Shopbell, P. 1996, ApJ, 471, L5
\reference{} Connolly, A. J., Csabai, I., Szalay, A. S., Koo, D. C., Kron,
             R. G., \& Munn, J. A. 1995, AJ, 110, 2655
\reference{} Connolly, A. J., Szalay, A. S., Dickinson, M. E., Subbarao, M., 
             U., \& Brunner, R. J. 1997, ApJ, 486, L11
\reference{} Cowie, L. L., Hu, E. M., \& Songaila, A. 1995a, Nature, 377, 603
\reference{} Cowie, L. L., Songaila, A., Hu, E. M. \& Cohen, J. G. 1996, AJ,
             112, 839
\reference{} Cowie, L. L., Hu, E. M., Songaila, A., \& Egami E. 1997, ApJ,
             481, L9 
\reference{} Davis, M., \& Peebles, P. J. E. 1983, ApJ, 267, 465
\reference{} Davis, M., Efstathiou, G., Frenk, C., \& White, S. D. M. 1985,
             ApJ, 292, 371
\reference{} Davis, M., Meiksin, A., Strauss, M. A., Da Costa, N. L., \& 
             Yahil, A. 1988, ApJ, 333, L9
\reference{} Efstathiou, G., Bernstein, G., Katz, N., Tyson, A. J., \&
             Guhathakurta, P. 1991, ApJ, 380, L47
\reference{} Efstathiou, G., 1995, MNRAS, 272, L25
\reference{} Eke, V. R., Cole, S., \& Frenk, C. S. 1996, MNRAS, 282, 263
\reference{} Giavalisco, M., Steidel, C. C., \& Macchetto 1996, ApJ, 470, 189
\reference{} Groth, E. J., \& Peebles, P. J. E. 1977, ApJ, 217, 385
\reference{} Governato, F., Baugh, C. M., Frenk, C. S., Lacey, C. G., Quinn, 
             T., \& Stadel, J. 1998, preprint
\reference{} Hamilton, A. J. 1988, ApJ, 331, L59
\reference{} Jing, Y. P., \& Suto, Y. 1998, ApJ, submitted, astro-ph/9710090
\reference{} Landy, S.D. \& Szalay, A.S., 1993, ApJ, 412, L64
\reference{} Landy, S.D. \& Szalay, A.S., \& Koo, D. C. 1996, ApJ, 460, 94
\reference{} Le F\`evre, O., Hudon, D., Lilly, S. J., Crampton, D., Hammer, F.,
             \& Tresse, L. 1996, ApJ, 461, 534
\reference{} Lilly, S., Tresse, L., Hammer, F., Crampton, D., \& Le F\`evre, 
             O. 1995, ApJ, 455, 108
\reference{} Ling, E. N., Barrow, J. D., Frenk, C. S. 1986, MNRAS, 223, L21
\reference{} Loveday, J., Maddox, S. J., Efstathiou, G., \& Peterson, B. A. 
             1995, ApJ, 442, 457
\reference{} Lowenthal, J. D., Koo, D. C., Guzman, R., Gallego, J., Phillips,
             A. C., Faber, S. M., Vogt, N. P., Illingworth, G. D., Gronwall, 
             C. 1997, ApJ, 481, 673
\reference{} Madau, P., 1995, ApJ, 441, 18
\reference{} Madau, P. Ferguson, H. C., Dickinson, M. E., Giavalisco, M.,
             Steidel, C. C., \& Fruchter, A. 1996, MNRAS, 283, 1388
\reference{} Madau, P., Pozzetti, L., \& Dickinson, M. E. 1997, ApJ, in press,
             astro-ph/9708220
\reference{} Mann, R. G., Peacock, J. A., \& Heavens, A. F. 1997, MNRAS, 
             in press, astro-ph/9708031
\reference{} Matarrese, S., Coles, P., Lucchin, F., \& Moscardini, L. 1997,
             MNRAS, 286, 115
\reference{} Mo, H. J., \& Fukugita, M. 1996, 467, L9
\reference{} Mo, H. J., \& White, S. D. M. 1996, MNRAS, 282, 347
\reference{} Oke, J. B., \& Gunn, J. E. 1983, ApJ, 266, 713
\reference{} Oke, J. B., Cohen, J. G., Carr, M., Cromer, J., Dingizian, A.,
             Harris, F. H., Labrecque, S., Lucinio, R., Schaal, W, Epps, H.,
             Miller, J. 1995, PASP, 107, 375
\reference{} Park, C., Vogeley, M. S., Geller, M., \& Huchra, J. P. 1994, ApJ,
             431, 569
\reference{} Peacock, J. A. 1997, MNRAS, 284, 885
\reference{} Peebles, P. J. E., 1980 ``The Large-Scale Structure of the 
             Universe'', Princeton University Press.
\reference{} Press, W. H., Flannery, B. P., Teukolsky, S. A., \& Vetterling, 
             W. T. 1992, ``Numerical Recipes'', Cambridge University Press
\reference{} Santiago, B. X., Da Costa, L. N. 1990, ApJ, 362, 386
\reference{} Schade, D., Lilly, S. J., Crampton, D., Hammer, F., Le F\`evre, 
             O., \& Tresse, L. 1995, ApJ, 451, L1
\reference{} Steidel, C.C., \& Hamilton, D. 1993, AJ, 105, 2017 (Paper II)
\reference{} Steidel, C.C., Pettini, M., \& Hamilton, D. 1995 (Paper III),
             AJ, 110, 2519
\reference{} Steidel, C. C., Giavalisco, M., Pettini, M., Dickinson, M., \& 
             Adelberger, K. 1996a, ApJ, 
\reference{} Steidel, C. C., Giavalisco, M., Dickinson, M., \& Adelberger,
             K. L. 1996b, AJ, 112, 352
\reference{} Steidel, C. C., Adelberger, K. L., Dickinson, M. E., Giavalisco, 
             M., Pettini, M., \& Kellogg, M. 1998, ApJ, in press,  
             astro-ph/9708125, Paper 1
\reference{} Tucker, D. L., Oemler, A., Jr., Kirshner, R. P., Lin, H.,
             Schectman, S. A., Landy, S. D., Schechter, P. L., Muller, V., 
             Gottlober, S., Einasto, J. 1997, MNRAS, 285, L5
\reference{} Valotto, C. A., \& Lambas, D. G. 1997, ApJ, 481, 594
\reference{} Weinberg, D., Katz, N., \& Hernquist, L. 1997, in ``Origins'', 
             eds. M. J. Shull, C. E. Woodward, \& H. Thronson, (ASP Conference 
             series), astro-ph/9708213
\reference{} White, S. M., Davis, M., Efstathiou, G., \& Frenk, C. S. 1987, 
             Nature, 330, 451
\reference{} White, S. M., Efstathiou, G., Frenk, C. S. 1993, MNRAS, 262, 1023
\end{references}
\end{document}